# Improving Family Co-Play Experiences through Family-Centered Design*

Zinan Zhang, Xinning Gui, & Yubo Kou

College of IST, Penn State, USA, zzinan@psu.edu, xinninggui@psu.edu, yubokou@psu.edu

Cooperative play, commonly abbreviated as co-play, represents a form of gaming where individuals engage in play together within a shared space. This concept of co-play has gained prominence alongside the advancements in gaming technologies such as Microsoft Xbox and Nintendo Wii, as well as the broader growth of the video game industry. Researchers have discovered the potential of these developments for fostering family-centered co-play experiences as co-play could enhance the relationship bond between family members [2]. Researchers have discovered a board spectrum of games facilitating family-centered co-play experiences, including genres from first-person shooter games, such as Call of Duty and Fortnite, to social simulation games like Animal Crossing, and extending to games in augmented reality, such as Pokemon Go and virtual reality, such as Roblox and Minecraft [8] The rise of extended reality (a collective term for virtual reality, augmented reality, and mixed reality) technology, such as Metaverse and Vision Pro may even open new avenues for co-play, including incorporating elements of both the physical and digital worlds. With the rise of extended virtual reality technology, researchers are increasingly interested in exploring the potential of co-play within these extended reality environments [4,7].

Despite the growing popularity and potential benefits of co-play with families in video games, not all co-play experiences are beneficial or playful. For example, problematic in-game behaviors of other players such as hate speech, harassment, and cyberbullying [12] can be disruptive to the co-play experiences of family members. Beyond problematic player behavior, our recent research has found several harmful designs in virtual world settings of video game, Roblox, that can disrupt player experiences with complex harm beyond merely harm in regular video game formats [6]. Roblox stands out as a gaming platform with over 40 million independent virtual worlds (termed 'experiences' by Roblox) and attracts more than 214 million monthly players, a significant majority of whom (85%) are younger than 18. These independent virtual worlds are primarily created by hobbyist creators rather than professional developers. Incorporating the definition of user-generated content, which represents users create and publishing their content on platforms [10], we called the virtual worlds on Roblox user-generated virtual worlds (UGVW), where virtual world is a type of content. Utilizing a grounded theory approach, we [6] found that UGVW designs can subtly manipulate players into certain behaviors, such as making purchases through prompts. In addition, these virtual worlds may feature unmoderated social designs that inappropriately engage players in specific actions within the virtual space. Players are also exposed to dynamic and emergent risks through exposure to unregulated content in real time. Lastly, problematic UGVW designs, such as those in storytelling, can encourage players' behaviors in virtual spaces aligned with extremist ideologies, values, and ideas.

The issue of harm within gaming platform like Roblox is particularly concerning given the platform's large base of child players. Previous literature on parental mediation in video gameplay identifies three main strategies that parents can employ to mitigate these risks: restrictive mediation, active mediation, and co-playing [9]. Restrictive

* We thank NSF No. 2326505 for supporting the work.

mediation involves setting limits on gaming activity, such as the duration of playtime or the types of games that can be played. Roblox supports this form of mediation through its parental controls [13], which allow parents to restrict their children's chat functions, spending amounts, and access to games based on age restrictions. These controls are designed to safeguard children from inappropriate content and interactions within the platform. Active mediation entails engaging in discussions about video game activities. This approach can help parents understand what their children are experiencing in these virtual worlds and provide guidance on how to navigate them safely. Through dialogue, parents can impart critical thinking skills and resilience strategies to help their children discern and cope with potential harm. Co-playing, where parents and children play together, serves as a direct form of engagement and supervision. Our recent research indicates that some parents choose to co-play with their children on Roblox, which may shield their children from certain harms targeted at them within the platform. By participating in the game alongside their children, parents can directly monitor interactions and intervene when necessary to prevent exposure to harmful content or behavior.

The presence of parental mediation in the form of co-playing may contribute to mitigating some risks associated with online gaming platforms like Roblox, particularly for its young audience. When parents participate alongside their children, they can directly address certain issues, such as in-game purchases, by discussing and setting boundaries to prevent excessive spending. This direct involvement allows for an immediate response to some of the potential harms targeted at children within these virtual spaces. Despite the benefits of parental presence, certain harms remain difficult to mitigate due to the nature of user-generated virtual worlds (UGVW) and the existing moderation mechanisms. For instance, the use of emotes to conduct inappropriate sexual interactions and the presence of unfiltered chat represent types [6] of harm that are not easily preventable by parents. This harm could negatively impact the co-play experiences with family members as they encounter the harm together in real time.

Moderation can be one approach to mitigate the harm that can interfere with the family co-play experiences in video games. Previous literature has discussed how to use content moderation and auto-moderation to moderate the harm in video games [1,5]. However, as the harm in virtual spaces is more complex compared to regular video game formats, our recent research highlights a critical gap in the effectiveness of current content moderation strategies on platforms like Roblox, which result in harm in UGVWs, which thus interferes the family co-play experiences [6]. These issues stem from the inherent limitations in the platform's content moderation strategies, which are primarily adapted from those used in traditional user-generated content platforms such as social media. Roblox's current governance strategies, including auto-moderation for harmful models, a reporting system, and chat filters, are designed to regulate these harms [13]. These strategies are typically used in user-generated content platforms and focus on managing static forms of content like text, videos, and images, which are fundamentally different from the dynamic and complex harm that occurs in virtual worlds. As such, the dynamic and complex harm evolves in real time and in ways that are not easily captured by traditional moderation tools. The disruption to co-play experiences due to these moderation challenges suggests that, although the presence of parents may improve the safety of the gaming environment compared to solitary play, it does not guarantee immunity from real-time harm, especially in virtual worlds. This reality underscores the need for UGVW platforms like Roblox to develop more advanced and nuanced content moderation strategies specifically tailored to the unique challenges of virtual worlds. Such strategies should aim to proactively identify and prevent potential harms in real-time, ensuring a safer and more enjoyable co-play experience for families.

While significant attention has been paid to enhancing the connection between parents and children through co-play [3,7,11] and protecting children from various harms associated with digital gaming environments through co-play [9], there is a notable gap in research concerning the specific harms that can disrupt co-play experiences and the strategies for mitigating the harm. This gap in the literature presents a critical area for investigation, particularly in the context of user-generated virtual worlds (UGVW) and their platforms. Thus, I propose a research question on how to design UGVWs and their hosting platforms in a manner that effectively minimizes the harm that could impair family centered co-play experiences. By focusing on these aspects, researchers and designers can develop UGVWs and their platforms that not only minimize harms but also enhance the quality of co-play experiences for families. This approach requires a concerted effort to merge child-centered safety research with family-centered design principles, ultimately creating a safer and more enriching digital playground for children and their parents to explore together.

To a larger extend, I would like to discuss in this workshop how to bridge the gap between child-centered safety research and family-centered design research. Addressing challenges of harm in family co-play experiences requires a nuanced understanding of safety issues within family-centered design. It's essential to recognize that children's safety needs during co-play may diverge significantly from what guardians anticipate [9]. This discrepancy suggests that the safety considerations for solitary play may not be directly applicable to co-play scenarios, raising important questions about whether and how safety protocols should be adjusted to accommodate the unique dynamics of co-play.